\documentclass[
aps,
prx,
twocolumn,
letterpaper,
superscriptaddress,
10pt
]{revtex4-2}

\usepackage{amssymb,amsthm,amsmath,amsfonts}
\usepackage{graphicx,enumerate,bbm,bm}
\usepackage[pdftex,dvipsnames]{xcolor}
\usepackage[colorlinks=true,urlcolor=blue,citecolor=blue,linkcolor=blue]{hyperref}
\usepackage{soul}

\begin{document}

\title{Quasiprobabilities from incomplete and overcomplete measurements}

\author{Jan Sperling}
    \affiliation{Paderborn University, Institute for Photonic Quantum Systems (PhoQS), Theoretical Quantum Science, Warburger Stra\ss{}e 100, 33098 Paderborn, Germany}

\author{Laura Ares}
    \affiliation{Paderborn University, Institute for Photonic Quantum Systems (PhoQS), Theoretical Quantum Science, Warburger Stra\ss{}e 100, 33098 Paderborn, Germany}

\author{Elizabeth Agudelo}
    \affiliation{TU Wien, Atominstitut \& Vienna Center for Quantum Science and Technology, Stadionallee 2, 1020 Vienna, Austria}

\date{\today}

\begin{abstract}
    We discuss the (re-)construction of quasiprobability representations from generic measurements, including noisy ones.
    Based on the measurement under study, quasiprobabilities and the associated concept of nonclassicality are introduced.
    A practical concern that we address is the treatment of informationally incomplete and overcomplete measurement scenarios, which can significantly alter the assessment of which states are deemed classical.
    Notions, such as Kirkwood--Dirac quasiprobabilities and $s$-parametrized quasiprobabilities in quantum optics, are generalized by our approach.
    Single-qubit systems are used to exemplify and to compare different measurement schemes, together with the resulting quasiprobabilities and set of nonclassical states.
\end{abstract}

\maketitle

\section{Introduction}

    Quantum information science leverages the fundamental principles of quantum mechanics to accomplish information processing and communication tasks that are beyond the reach of classical physics \cite{2010NielsenChuang, 2009HorodeckiHorodeckiHorodeckiHorodecki}. 
    Over recent decades, an extensive range of experiments has demonstrated the feasibility of diverse quantum protocols, which require different techniques, ranging from the preparation and manipulation of simple qubits to the control of complex entangled systems. 
    A central role for these protocols and for quantum theory itself is played by the quantum state, which encodes the full physical description of a system.
    Understanding and determining this state is therefore essential for evaluating the evolution of the system, identifying distinct quantum features, and quantifying its usefulness as a resource for quantum technologies.

    One quantum state, however, can be represented in multiple equivalent forms.
    Quantum state representations include wavefunctions, Hilbert-space vectors, density operators, etc.
    Phase-space descriptions are one particularly interesting approach, where quantum states are represented through joint distributions defined over a pair of conjugate variables 
    \cite{Glauber_1963,Sudarshan_1963}.
    The earliest and most influential of these is the phase-space distribution was introduced by Wigner \cite{Wigner_1932} and was originally conceived as a probability representation of wavefunctions.
    Soon after, it was fully recognized as a quantum phase-space equivalent to the probability density functions \cite{Moyal_1949}. 
    This distribution, now known as the Wigner function, plays a central role in continuous-variable quantum theory by providing a direct analog to classical phase-space descriptions. 
    However, unlike true probabilities, it can take on negative values in certain regimes, thus earning the designation of a quasiprobability distribution.
    The concept of phase-space representations has been further generalized to the family of $s$-parametrized quasiprobabilities \cite{CahillGlauber_1969b}, providing a continuous interpolation between different types of phase-space functions and allowing a unified description of quantum states.
    This includes the $P$ function ($s=1$) \cite{Glauber_1963,Sudarshan_1963}, Wigner function ($s=0$) \cite{Wigner_1932}, and Husimi $Q$ function ($s=-1$) \cite{Husimi_1940} as special cases.
    For instance, in quantum optics, the impossibility of interpreting the Glauber-Sudarshan $P$-function as a classical probability density for certain states defines the concept of nonclassicality \cite{TitulaerGlauber_1965, 1986Mandel}.
    Negative values in quasiprobabilities are the signature of quantumness; see, e.g., Ref. \cite{2016Sperling} for an overview.
    Classical modeling is possible exclusively in cases where joint probability distributions capture every aspect of a system’s preparation, transformation, and observation \cite{Dirac_1926,Wigner_1932,Dirac_1942,Dirac_1945,Cohen_1966b,Hudson_1974,SrinivasWolf_1975,Griffiths_1984,Hartle_2004,MariEisert_2012,Rahimi-KeshariRalphCaves_2016,Allahverdyan_2015}. 
    Identifying the situations in which no classical counterpart exists remains a long-standing difficulty.

    Quasiprobability representations need not be real-valued as the observables defining the joint distribution may not be mutually compatible \cite{SprietLangrenezBrummelhuisDeBievre_2025}.
    The Kirkwood–-Dirac (KD) distribution \cite{Kirkwood_1933, Dirac_1945,Rihaczek_1968}, for instance, can represent quantum states with respect to arbitrary pairs of observables, making the complex values of the distribution also a signature of nonclassicality.
    The KD is closely linked to weak values \cite{DresselMalikMiattoEtAl_2014}, which can be interpreted as conditional expectation values of non-Hermitian operators. 
    This connection highlights how, in contrast to the $s$-parametrized quasiprobabilities, the KD formalism provides a phase-space representation in which the statistics of potentially incompatible measurements naturally give rise to complex weak values.
    A noteworthy relative to the KD distribution is the Terletsky--Margenau--Hill distribution \cite{Terletsky_1937,MargenauHill_1961,Johansen_2004,JohansenLuis_2004}.
        
    The concept of nonclassicality defined by the negative values of quasiprobability representations is strongly related to the chosen classical reference, i.e., to the convex set of states or observables chosen because their share some ``classical'' property central to the present analysis.
    Thus, quasiprobability representations constitute a tool to define arbitrary notions of nonclassicality, relative to the physical feature under study \cite{2018SperlingWalmsley2}.
    For this reason, over the years, quasiprobability methods have become a cornerstone for identifying genuinely quantum phenomena---such as interference, noncommutativity, contextuality, entanglement, and other quantum correlations---and distinguishing them from their classical analogues, even in hybrid systems \cite{Vogel_2000, Spekkens_2008,  SperlingVogel_2009, FerraroParis_2012, AgudeloSperlingVogel_2013, JeongZavattaKangEtAl_2014, MorinHuangLiuEtAl_2014, AgudeloSperlingCostanzoEtAl_2017, HuangJeannicMorinEtAl_2019, GarciaAresLuis_2019, LuisAres_2020,MasaAresLuis_2020, DeBievre_2021, Arvidsson-ShukurDroriHalpern_2021, BoothChabaudEmeriau_2022, DeBievre_2023, KohnkeAgudeloSchunemannSchlettweinVogelSperlingHage_2021, SperlingGiananiBarbieriAgudelo_2023, SchmidBaldijaoYingEtAl_2024, AresPrasannanAgudeloEtAl_2024, LiuGaoFadelEtAl_2025, ZawGuoQeEtAl_2025}.
    They also serve as indispensable tools for identifying and characterizing quantum states in optical systems and beyond, which has been recently reviewed and extended \cite{SperlingVogel_2020,Arvidsson-ShukurBraaschDeBievreEtAl_2024}.
   
    Regardless of the representation, the information about the state of the system is determined in experiments.
    A full state reconstruction is usually attempted through quantum state tomography, a family of experimental and computational procedures designed to reconstruct the quantum state from statistical data obtained through various measurements \cite{1997Leonhardt, 2001JamesKwiatMunroWhite, 2004ParisRehacek, AgudeloSperlingVogelEtAl_2015}.
    To fully and uniquely identify the state, the chosen measurement operators must constitute a tomographically complete set, meaning that they span an operator basis over the system’s Hilbert space and thus provide access to the full information content of the state and that they do not include redundancies for uniqueness.
    However, implementing a tomographically complete measurement set is often resource-intensive, requiring a large number of carefully chosen measurement settings and extensive data collection.
    In realistic experimental scenarios, such demands can quickly become prohibitive, especially for systems of high dimensionality and multiple subsystems. 
    For example, a tomography in a Hilbert space of dimension $d$ requires performing $d^2$  measurements to account for each density-operator basis element. 
    This challenge naturally gives rise to the central question:
    What claims about the quantum state can still be made when the available measurements are not informationally complete; that is, when are the experimental data insufficient or redundant for a full and unique reconstruction?

    In this work, we construct quasiprobability distributions for arbitrary measurements, addressing the representation of the state in terms of the available set of measurements.
    In this context, the definition of nonclassicality is relative to the set of measurements under consideration.
    Using pseudo-inversion strategies, we extend our framework to detection scenarios which are informationally incomplete, i.e., insufficient to reconstruct the full quantum state, and informationally overcomplete, i.e., the reconstruction of the quasiprobability is not unique.
    Known representations, such as KD and $s$-parametrized phase-space functions, are shown to be a special case of our universally applicable approach.
    Proof-of-concept examples of distinct detection scenarios of a qubit are employed to highlight the impact of noise, incompleteness, and overcompleteness when assessing negativities in the constructed quasiprobabilities.

    The remainder of the manuscript is structured as follows:
    In Sec. \ref{sec:Methodology}, the general framework to measurement-based quasiprobabilities for incomplete and overcomplete measurements is formulated.
    Single-qubit examples in Sec. \ref{sec:Applications} show the impact of the here-considered measurement imperfections for a system with importance for quantum information science.
    We summarize our findings in Sec. \ref{sec:Conclusion}.

\section{Methodology}
\label{sec:Methodology}

    In this section, we formulate the general approach to measurement-based quasiprobabilities from informationally incomplete and overcomplete measurements.
    Beginning with the informationally complete case, partial pseudo-inversions in form of convolutions are used to address a lack of informational completeness in the non-ideal detection scenario.
    Relations to seminal quasiprobabilities, such as KD and $s$-parametrized phase-space representations, are discussed.

\subsection{Quasiprobabilities from measurements}

    Consider a measurement defined by a set of operators $\{\hat\Pi_k\}_k$, where $k$ may denote either a single index or a multi-index, e.g., $k = (k_1, k_2, \ldots)$, as will be the case later in this section.
    The measurement outcomes are given by the distribution
    \begin{equation}
        Q(k)=\mathrm{tr}(\hat\Pi_k^\dag\hat\rho),
    \end{equation}
    being the Hilbert--Schmidt inner product between the measurement operators and the state, and where we allow for non-hermitian operators, explicitly allowing for weak measurements.
    The vector of outcomes, $\vec Q=[Q(k)]_k$, presents all the information known about the quantum state $\hat\rho$ we can infer from the measurement.
    Further, by definition, a measurement is said to be informationally complete if a state, $\hat \rho$, is fully and uniquely described through the outcomes $\vec Q$.
    
    Now, we can introduce a metric tensor $\boldsymbol g$ via Hilbert--Schmidt inner product of measurement operators.
    The entries of such a metric tensor read
    \begin{equation}
        g_{k,l}=\mathrm{tr}(\hat\Pi_k^\dag\hat\Pi_l).
    \end{equation}
    Introducing this notion is convenient because it allows us to determine the dual basis as
    \begin{equation}
        \hat\rho_k=\sum_l g^{-1}_{l,k}\hat\Pi_l,
    \end{equation}
    which obeys an orthonormality relation given by the Kronecker symbol $\delta_{k,l}=\mathrm{tr}(\hat\Pi^\dag_k\hat\rho_l)$.
    Note that, for now, we assume that the inverse metric tensor $\boldsymbol{g}^{-1}$ exists.
    Further note that $\hat\rho_j$ as defined above is not necessarily a positive semidefinite operator, $\hat\rho_j\ngeq0$, thus not a proper density operator, yielding the notion of a quasi-state \cite{2018SperlingWalmsley}, and which is studied later in greater detail.

    Through the notion of the dual basis, it is straightforward to see that we can expand the state under study as
    \begin{equation}
        \hat\rho=\sum_k Q(k)\hat\rho_k
        =\sum_{l} P(l)\hat\Pi_l,
    \end{equation}
    with $P(l)=\sum_{k} g^{-1}_{l,k}Q(k)$.
    Analogously to $\vec Q$, we may introduce the vector $\vec P=[P(k)]_k$ for convenience.
    In a vector-valued representation, we can equivalently write the above relation as
    \begin{equation}
        \vec P=\boldsymbol{g}^{-1}\vec Q
        \quad\text{and}\quad
        \vec Q=\boldsymbol{g}\vec P.
    \end{equation}
    The elements of $\vec P$ define the sought-after quasiprobability, which enables an expansion of the quantum state in terms of the measurement operators.
    Nonclassicality, manifested as negative components of $\vec{P}$, reflects the impossibility of expressing the state as a classical mixture of POVM elements, thereby signaling the necessity of quantum superpositions.

\paragraph*{Example.}

    As a first example, we take two orthonormal basis $\{|a_k\rangle\}_k$ and $\{|b_l\rangle\}_l$ and define
    \begin{equation}
        \hat\Pi_{(k,l)}=\frac{|a_k\rangle\langle b_l|}{\langle b_l|a_k\rangle},
    \end{equation}
    assuming $\langle b_l|a_k\rangle\neq0$.
    The above description presents a general weak-measurement scenario, with outcomes $Q(k,l)=\langle b_l|\hat\rho|a_k\rangle/\langle b_l|a_k\rangle$.
    The entries of the metric tensor form a diagonal matrix,
    \begin{equation}
        g_{(k,l),(k',l')}
        =\frac{\delta_{k,k'}\delta_{l,l'}}{|\langle a_k|b_l\rangle|^2},
    \end{equation}
    whose inverse can be readily determined.
    Thus, the elements of the quasiprobability read
    \begin{equation}
        P(k,l)
        =|\langle a_k|b_l\rangle|^2
        Q(k,l)
        =
        \langle a_k|b_l\rangle \langle b_l|\hat\rho|a_k\rangle,
    \end{equation}
    which is the most common definition of the KD distribution described as a special case of our general treatment;
    see Ref. \cite{Arvidsson-ShukurBraaschDeBievreEtAl_2024} for a broad overview.

\subsection{Convolution and $\sigma$-parametrization}

    Rather than a full deconvolution, $\vec P=\boldsymbol{g}^{-1}\vec Q$, one can study partial deconvolutions, where
    \begin{equation}
        \vec P_\sigma = \boldsymbol{g}^{-\sigma} \vec Q,
        \quad\text{likewise}\quad
        \vec P_\sigma = \boldsymbol{g}^{1-\sigma} \vec P,
    \end{equation}
    with $\boldsymbol{g}^0=\mathbbm 1$ being defined as the identity and $0\leq \sigma\leq 1$, also resulting in $\vec Q=\vec P_0$ and $\vec P=\vec P_1$.
    Similarly, the dual basis can be generalized by introducing the following operators:
    \begin{equation}
        \hat \Delta_\sigma(j)
        =
        \sum_{l} g^{-\sigma}_{l,j}\hat \Pi_l
    \end{equation}
    such that $\hat \Delta_1(j)=\hat\rho_j$ and $\hat\Delta_0(j)=\hat\Pi_j$.
    Also note that $\sigma<0$ can be used as noise model for measurements, as described later through examples.
    
    Importantly, we can more generally expand the state under study as
    \begin{equation}
        \hat\rho=\sum_l P_\sigma(l)\hat\Delta_{1-\sigma}(l).
    \end{equation}
    With this expansion, we define $\hat\rho$ as $\sigma$-classical if $\vec P_\sigma\geq0$.
    In other word, a $\sigma$-classical state obeys
    \begin{equation}
        \forall k: P_\sigma(k)\geq0.
    \end{equation}
    Otherwise, we say $\hat\rho$ is $\sigma$-nonclassical.
    It is also noteworthy that if $\hat\Delta_{0}(j)\geq0$ holds true for all $j$, it defines a positive operator-valued measure (POVM), and its dual---given by the elements $\hat\Delta_{1}(j)$---has been dubbed a contravariant operator-valued measure \cite{2018KovalenkoSperlingVogelSemenov}.

\paragraph*{Example.}

    Suppose a continuum of measurement operators, $\hat\Pi_\alpha=|\alpha\rangle\langle \alpha|/\pi$, where $|\alpha\rangle$ denotes coherent states of a quantized harmonic oscillator, with $\alpha\in\mathbb C$.
    In quantum optics, such a measurement is realized through eight-port homodyning.
    The convolution kernel reads
    \begin{equation}
        g_{\alpha,\beta}
        =
        \mathrm{tr}(\hat\Pi_\alpha^\dag\hat\Pi_\beta)
        =
        \frac{1}{\pi}\frac{\exp\left(-|\alpha-\beta|^2\right)}{\pi},
    \end{equation}
    which is a Gaussian distribution---up to the factor $\pi^{-1}$.
    The measurement outcomes define the Husimi $Q$ function, $Q(\beta)=\langle\beta|\hat\rho|\beta\rangle/\pi=P(\beta;-1)$, where $P(\beta; s)$ denotes the $s$-parametrized quasiprobabilities known from quantum optics \cite{CahillGlauber_1969b}.

    A deconvolution of $Q$ yields an expression proportional to the seminal Glauber--Sudarshan $P$ function,
    \begin{equation}
        P(\alpha)
        =\int d^2\beta\,g^{-1}_{\alpha,\beta}\, Q(\beta)
        =\pi P(\alpha;+1).
    \end{equation}
    In general, we can find a relation between $P_\sigma$ and $s$-parametrized quasiprobabilities as
    \begin{equation}
    \begin{aligned}
        P_\sigma(\alpha)
        ={}&
        \pi^{-(1-\sigma)}
        \int d^2\beta \frac{\exp\left(-\frac{|\alpha-\beta|^2}{1-\sigma}\right)}{\pi(1-\sigma)}
        P(\beta)
        \\
        ={}&
        \pi^\sigma P\left(\alpha;2\sigma-1\right).
    \end{aligned}
    \end{equation}
    The identity $\sigma=(1+s)/2$ allows us to find all $s$-parametrized quasiprobabilities as a special scenario of our general treatment.
    In this context, we also observe that the value $\sigma=1/2$ is obtained from $s=0$, which defines the Wigner function, i.e.,
    \begin{equation}
        W(\alpha)=P(\alpha;0)=\frac{P_{1/2}(\alpha)}{\sqrt\pi}.
    \end{equation}
    Finally, we can, for example, write
    \begin{equation}
        \hat\rho
        =\int d^2\alpha\, P_1(\alpha)\,\hat\Delta_0(\alpha)
        =\int d^2\alpha\,P(\alpha;1)\,|\alpha\rangle\langle\alpha|,
    \end{equation}
    which is the well-established state's Glauber-Sudarshan representation in continuous variables.
    
\subsection{Overcomplete and incomplete measurements}

    Within the framework introduced and reviewed above, a crucial assumption is the informational completeness of the measurement $\{\hat\Pi_k\}_k$.
    However, this is rarely true in practice, where too few and redundant information is commonly measured.
    In the remainder of this work, we therefore address this concern with a careful and rigorous generalization of the aforementioned concepts, together with several applications.
    
    An informationally \textit{incomplete} measurement describes a measurement for which $\vec Q$ does not allow to infer all information about the quantum state under study.
    An informationally \textit{overcomplete} measurement is a scenario in which parts of $\vec Q$ are already sufficient to determine the full quantum state.

    In a mathematical formulation, the above explanations translate to a state expansion
    \begin{equation}
        \hat\rho=\sum_{k}P(k)\hat\Pi_k+\hat\nu.
    \end{equation}
    Therein, we have an orthogonal complement $\hat\nu$ to account for residual components that are invisible to an incomplete measurement;
    that is, we have
    \begin{equation}
        \forall l:
        \mathrm{tr}(\hat\Pi_l^\dag\hat\nu)=0.
    \end{equation}

    Overcompleteness is captured by a non-invertible $\boldsymbol{g}$.
    This means that we have non-unique $\vec P\neq\vec P'$ such that the measurement outcomes are not affected, $\boldsymbol{g}\vec P=\vec Q=\boldsymbol{g}\vec P'$.
    Likewise, we can equate
    \begin{equation}
        \vec P=\vec P'+\vec N,
        \quad\text{with}\quad
        \boldsymbol{g}\vec N=0
    \end{equation}
    defining a nullspace element $\vec N$.
    Note that $\vec P'$ can be obtained from $\vec P'=\boldsymbol{g}^{-1}\vec Q$, where $\boldsymbol{g}^{-1}$ now denoting the pseudo-inverse---rather than the proper inverse.
    Generalizations to $\sigma$-parametrized quasiprobabilities trivially follow from the earlier sections.
    
    In the remainder of this work, we investigate proof-of-concept examples that apply the methodology formulated above.
    We begin with an informationally complete measurement and $\sigma$-nonclassicality to describe noisy measurement.
    Still, studying the significant impact of informationally incomplete and overcomplete measurements is the main goal of most of the following applications, specifically focusing on the aforementioned notion of $\sigma$-nonclassical states.

\section{Applications}
\label{sec:Applications}

    In the following, we explore a qubit as a fundamental quantum system with a wide range of applications and realizations, such as the polarization of a photon.
    See, e.g., Ref. \cite{GiraudBraunBraun_2008} for an introduction.
    The measurements we are going to study define positive operator-valued measures (POVMs).
    That is, we have $\hat\Pi_k=\hat\Pi_k^\dag\geq0$, for all $k$, and $\sum_k\hat\Pi_k=\mathbbm 1_2$, where $\mathbbm 1_2$ denotes the $2\times 2$ identity.
    In addition, a generic quantum state may be expanded in terms of Pauli matrices, $\{\hat\sigma_x,\hat\sigma_y,\hat\sigma_z\}$, as
    \begin{equation}
    \begin{aligned}
        \hat\rho
        ={}&
        \frac{1}{2}\left(
            \mathbbm 1_2
            +x\hat\sigma_x
            +y\hat\sigma_y
            +z\hat\sigma_z
        \right)
        \\
        ={}&
        \frac{1}{2}
        \begin{bmatrix}
            1+z & x-i y
            \\
            x+i y & 1-z
        \end{bmatrix}
        ,
    \end{aligned}
    \end{equation}
    where $x^2+y^2+z^2\leq 1$.
    The following examples cover all essential cases, which include noisy, incomplete, and overcomplete measurement scenarios.

\subsection{Noisy POVM}

    To define the ideal measurement in the first step, we begin with the normalized vectors
    \begin{equation}
        |\psi_3\rangle=|0\rangle
        \quad\text{and}\quad
        |\psi_j\rangle=\frac{|0\rangle+\sqrt2\omega^j|1\rangle}{\sqrt3},
    \end{equation}
    for $j\in\{0,1,2\}$ and $\omega=\exp(2\pi i/3)$.
    Those states span a tetrahedron on the Bloch sphere.
    The initially noisless POVM is defined through rank-one operators as
    \begin{equation}
        \hat\Pi_j=\frac{1}{2}|\psi_j\rangle\langle \psi_j|,
    \end{equation}
    where $j\in\{0,1,2,3\}$, obeying positive semi-definiteness, $\hat\Pi_j\geq0$, and the completeness relation, $\sum_{j=0}^3\hat\Pi_j=\mathbbm 1_2$.
    
    One can directly compute the metric tensor as
    \begin{gather}
        g_{j,k}
        ={}
        \frac{1}{12}+\frac{1}{6}\delta_{j,k}\nonumber
        \\
        \boldsymbol{g}
        ={}
        \frac{1}{6}\mathbbm 1_4+\frac{1}{12}\vec n_4\vec n_4^\mathrm{T}
        ={}
        \frac{1}{12}
        \begin{bmatrix}
            3 & 1 & 1 & 1
            \\
            1 & 3 & 1 & 1
            \\
            1 & 1 & 3 & 1
            \\
            1 & 1 & 1 & 3
        \end{bmatrix}
        ,
    \end{gather}
    where $\vec n_4$ denotes a four-dimensional vector of ones.
    The spectral decomposition of $\boldsymbol{g}$ allows us to find the $a$\textsuperscript{th} power of the metric tensor,
    \begin{equation}
    \begin{aligned}
        \boldsymbol{g}
        ={}&
        \frac{1}{6}\left(
            \mathbbm 1_4
            -\frac{\vec n_4\vec n_4^\mathrm{T}}{4}
        \right)
        +
        \frac{1}{2}
        \frac{\vec n_4\vec n_4^\mathrm{T}}{4}
        \\
        \rightarrow\quad
        \boldsymbol{g}^a
        ={}&
        6^{-a}\mathbbm 1_4
        +
        \frac{2^{-a}-6^{-a}}{4}\vec n_4\vec n_4^\mathrm{T}.
    \end{aligned}
    \end{equation}
    For instance, we can compute the $\sigma$-parametrization of the basis operators as
    \begin{equation}
    \begin{aligned}
        {}&
        \hat\Delta_{1-\sigma}(l)
        =
        \sum_{k=0}^3\left(
            6^{1-\sigma}\delta_{k,l}
            +\frac{2^{1-\sigma}-6^{1-\sigma}}{4}
        \right)\hat\Pi_k
        \\
        ={}&
        \frac{6^{1-\sigma}+2^{1-\sigma}}{4}
        |\psi_l\rangle\langle\psi_l|
        +
        \frac{2^{1-\sigma}-6^{1-\sigma}}{4}\left(
            \mathbbm 1_2-|\psi_l\rangle\langle\psi_l|
        \right),
    \end{aligned}
    \end{equation}
    which is a full-rank operator.
    Particularly, we describe a noisy measurement for $1-\sigma<0$ (i.e., $\sigma>1$), and we have a quasistate for $1-\sigma>0$ (i.e., $\sigma<1$) because we have $\hat\Delta_{1-\sigma}(l)\ngeq0$ in this case.
    
    Furthermore, the outcome vector $\vec Q$ can be related to the Pauli-matrix expansion of the state as follows:
    \begin{equation}
    \begin{aligned}
        \vec Q
        ={}&
        [\mathrm{tr}(\hat\Pi_k^\dag\hat\rho)]_{j\in\{0,1,2,3\}}
        \\
        ={}&
        \frac{1}{12}
        \begin{bmatrix}
            3 & 3 & 0 & 0
            \\
            3 & -1 & 2\sqrt2 & 0
            \\
            3 & -1 & -\sqrt2 & \sqrt6
            \\
            3 & -1 & -\sqrt2 & -\sqrt6
        \end{bmatrix}
        \begin{bmatrix}
            1 \\ z \\ x \\ y
        \end{bmatrix}.
    \end{aligned}
    \end{equation}
    The matrix connecting Bloch-sphere coordinates with the measurment outcomes is bijective since the measurement is informationally complete.
    Next, the $\sigma$-parametrized quasiprobabilities can be computed as
    \begin{equation}
    \begin{aligned}
        {}&
        \vec P_\sigma
        =
        \boldsymbol{g}^{-\sigma}\vec Q
        \\
        ={}&
        \frac{6^\sigma}{12}\left(
            3^{1{-}\sigma}
            \begin{bmatrix}
                1 \\ 1 \\ 1 \\ 1
            \end{bmatrix}
            {+}
            z
            \begin{bmatrix}
                3 \\ -1 \\ -1 \\ -1
            \end{bmatrix}
            {+}
            \sqrt{2}x
            \begin{bmatrix}
                0 \\ 2 \\ -1 \\ -1
            \end{bmatrix}
            {+}
            \sqrt{6}y
            \begin{bmatrix}
                0 \\ 0 \\ 1 \\ -1
            \end{bmatrix}
        \right).
    \end{aligned}
    \end{equation}
    Restricting the components of $\vec P_\sigma$ to non-negative numbers allows us to find the set of $\sigma$-classical states as the ones that obey
    \begin{equation}
    \begin{aligned}
          -\frac{1}{3^\sigma}\leq z \leq 3^{1-\sigma}+2\sqrt2 x
        \text{ and }
        |y|\leq \frac{3^{1-\sigma}-z-\sqrt2 x}{\sqrt6}.
    \end{aligned}
    \end{equation}

\begin{figure*}
	\includegraphics[width=.17\textwidth]{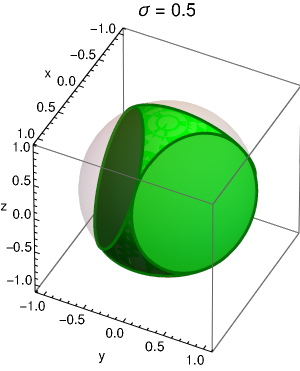}
	\hfill
	\includegraphics[width=.17\textwidth]{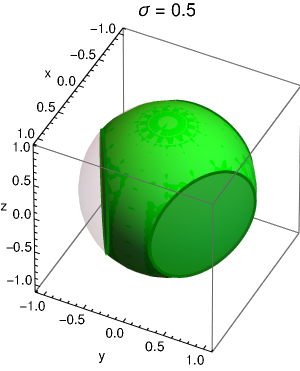}
	\hfill
	\includegraphics[width=.17\textwidth]{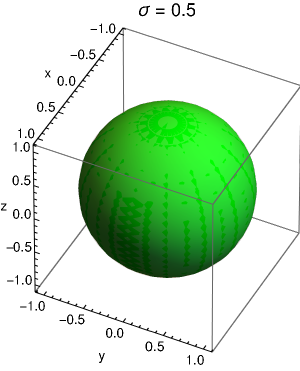}
	\hfill
	\includegraphics[width=.17\textwidth]{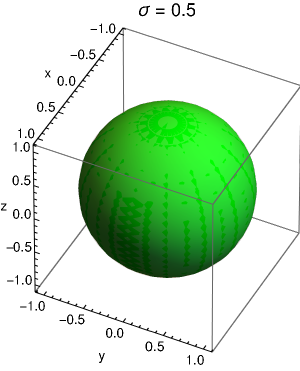}
	\\[2ex]
	\includegraphics[width=.17\textwidth]{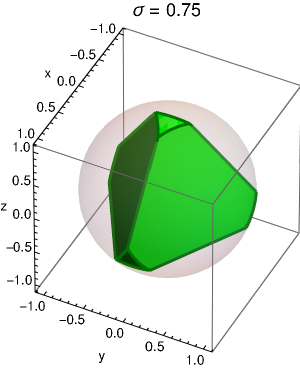}
	\hfill
	\includegraphics[width=.17\textwidth]{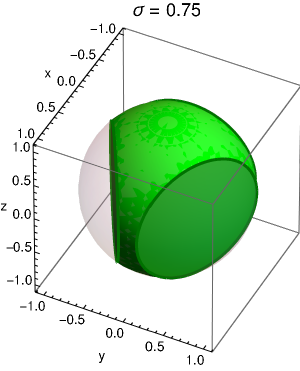}
	\hfill
	\includegraphics[width=.17\textwidth]{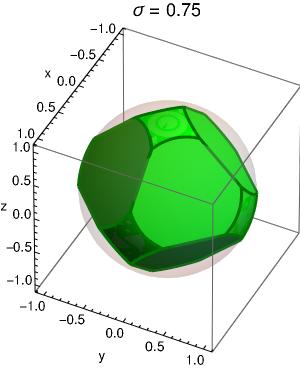}
	\hfill
	\includegraphics[width=.17\textwidth]{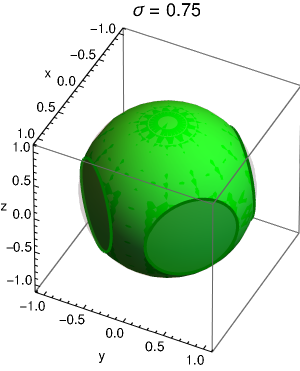}
	\\[2ex]
	\includegraphics[width=.17\textwidth]{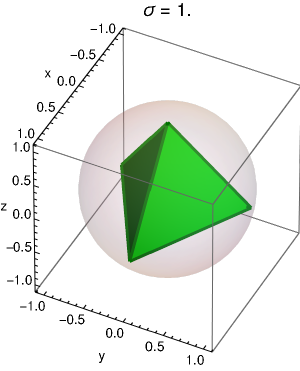}
	\hfill
	\includegraphics[width=.17\textwidth]{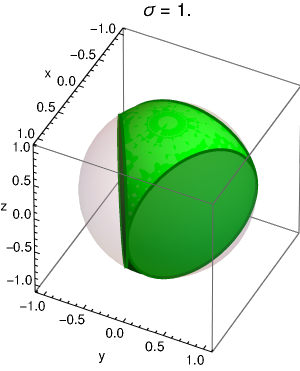}
	\hfill
	\includegraphics[width=.17\textwidth]{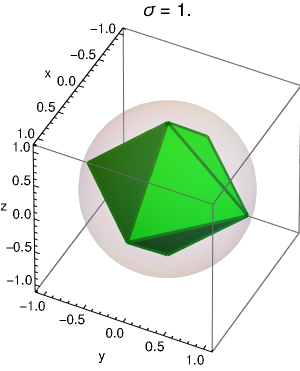}
	\hfill
	\includegraphics[width=.17\textwidth]{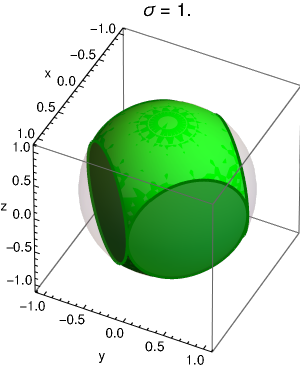}
	\\[2ex]
	\includegraphics[width=.17\textwidth]{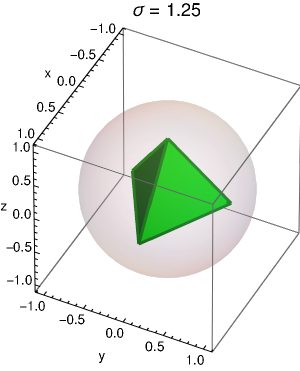}
	\hfill
	\includegraphics[width=.17\textwidth]{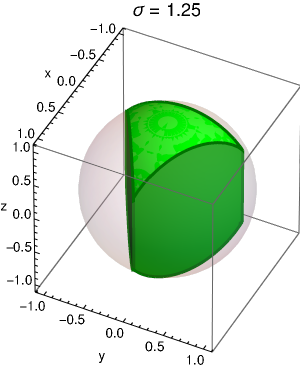}
	\hfill
	\includegraphics[width=.17\textwidth]{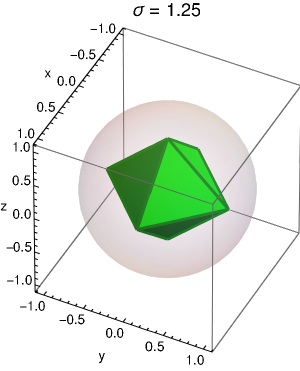}
	\hfill
	\includegraphics[width=.17\textwidth]{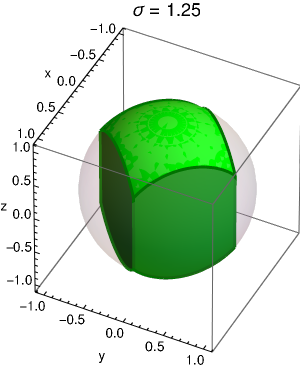}
	\caption{%
            Qubit measurements.
            Classical states with $\vec{P}_\sigma \geq 0$ are shown as green convex regions embedded in the Bloch sphere.
            The first column illustrates an informationally complete POVM; the second, an incomplete POVM lacking information about the $z$ axis.
            The third column depicts an overcomplete POVM, while the last shows a measurement set that is both informationally incomplete and overcomplete.
            The first row, $\sigma=0.5$, resembles the Wigner function for the measurement-based quasiprobabilities $\vec P_\sigma$.
            The second row, $\sigma=.75$, is closer to the analog of the Glauber--Sudarshan distribution, which itself is depicted in third row and is given by $\sigma=1$.
            Additional noise in the measurement can be modeled with $\sigma>1$, which is shown in the last row for $\sigma=1.25$.
            States outside the green set are $\sigma$-nonclassical, $\vec P_\sigma\ngeq0$, for the selected $\sigma$ parameter and POVM under consideration.
	}\label{fig:allPlots}
\end{figure*}

    In Fig. \ref{fig:allPlots}, the first column of plots depicts the classical states---indicated as green, convex sets---for different $\sigma$ values and POVM currently under study.
    Nonclassical states in and on the Bloch sphere with $\vec P_\sigma\ngeq 0$ are outside this set.
    The third row pertains to the case $\vec P$, i.e., $\sigma=1$, and highlights the tetrahedron spanned by the POVM under study.
    The first row resembles the Wigner function, $\sigma=1/2$, which is less sensitive as it finds fewer states to be nonclassical.
    Geometrically speaking, the classical set in this case is the intersect of the Bloch ball with an up-scaled tetrahedron.
    An intermediate value, $0<\sigma<1$, is depicted in the second row for comparison.
    Note that, for $\sigma=0$, all members of Bloch ball are found to be classical, $\vec Q\geq0$, which is not depicted.
    The fourth row, where $\sigma>1$, models a noisy POVM, characterized through a down-scaled tetrahedron.
    That is, fewer states can be expressed as convex mixtures of the now noisy POVM elements.

\subsection{Informationally incomplete POVM}

    For giving the essential example of an informationally incomplete measurement, we consider states that span a triangle in the equatorial plane of the Bloch sphere,
    \begin{equation}
        |\psi_{j}\rangle=\frac{|0\rangle+\omega^j|1\rangle}{\sqrt2},
        \quad\text{for }
        j\in\{0,1,2\}.
    \end{equation}
    Thus, these states do not provide information about the $z$ direction.
    The POVM that can be constructed is
    \begin{equation}
        \hat\Pi_{j}=\frac{2}{3}|\psi_j\rangle\langle\psi_j|,
    \end{equation}
    for which $\hat\Pi_j\geq0$ and $\sum_{j=0}^2\hat\Pi_j=\mathbbm 1_2$ can be verified straightforwardly.
    
    As seen from the previous example, the key ingredients to assess quasiprobabilities are the spectral decomposition of the metric tensor $g$ and the relations of outcome vector $\vec Q$ to the states' Pauli-matrix expansion.
    The former is given by
    \begin{equation}
    \begin{aligned}
        \boldsymbol{g}
        ={}&
        \left[\frac{1}{9}+\frac{1}{3}\delta_{j,k}\right]_{j,k\in\{0,1,2\}}
        \\
        ={}&
        \frac{2}{3}\frac{\vec n_3\vec n_3^\mathrm{T}}{3}
        +\frac{1}{3}\left(
            \mathbbm 1_3-\frac{\vec n_3\vec n_3^\mathrm{T}}{3}
        \right)
        =
        \frac{1}{9}
        \begin{bmatrix}
            4 & 1 & 1
            \\
            1 & 4 & 1
            \\
            1 & 1 & 4
        \end{bmatrix}
        ,
    \end{aligned}
    \end{equation}
    and the latter takes the form
    \begin{equation}
        \vec Q
        =
        \frac{1}{6}
        \begin{bmatrix}
            2 & 0 & 2 & 0
            \\
            2 & 0 & -1 & \sqrt3
            \\
            2 & 0 & -1 & -\sqrt3
        \end{bmatrix}
        \begin{bmatrix}
            1 \\ z \\ x \\ y
        \end{bmatrix}.
    \end{equation}
    The matrix involved is not injective.
    Specifically, we have $\hat\nu=z\hat\sigma_z/2$ as the free orthogonal complement to the incomplete measurement under study.
    That is, the $z$ component of the state does not contribute towards the classical-quantum discrimination for the informationally incomplete case studied here.
    
    As done in the previous example, we can now compute
    the quasiprobabilities, $\vec P_\sigma=\boldsymbol{g}^{-\sigma}\vec Q$, which read
    \begin{equation}
        \vec P_\sigma
        =
        \frac{3^\sigma}{6}\left(
            2^{1{-}\sigma}
            \begin{bmatrix}
                1 \\ 1 \\ 1
            \end{bmatrix}
            +
            x
            \begin{bmatrix}
                2 \\ -1 \\ -1
            \end{bmatrix}
            +\sqrt3 y
            \begin{bmatrix}
                0 \\ 1 \\ -1
            \end{bmatrix}
        \right).
    \end{equation}
    From this, we conclude that $\sigma$-classical states for the currently studied POVM obey
    \begin{equation}
        x\geq -\frac{1}{2^\sigma}
        \text{ and }
        |y|\leq\frac{2^{1-\sigma}-x}{\sqrt 3}.
    \end{equation}
    
    In Fig. \ref{fig:allPlots}, the second column highlights the set of states with $\vec P_\sigma\geq0$ for the here-investigated incomplete measurement, probing different $\sigma$ values in the rows.
    A Wigner-like case ($\sigma=1/2$, first row), a Glauber--Sudarshan-like case ($\sigma=1$, third row), intermediate case ($0<\sigma<1$, second row), and a noisy-measurement case ($\sigma>0$, fourth row) are depicted.
    Since no information is collected about the $z$-direction, we can observe from the plots that the property of being classical or non-classical does not depend on the $z$ direction.
    In geometric terms, the members that are classical are found in the intersect of the Bloch ball and a scaled---depending on $\sigma$---prism with a triangular base that reflects the properties of the POVM under consideration.
    
\subsection{Informationally overcomplete POVM}

    For the informationally overcomplete case, we consider an collection of vectors with an octahedron configuration on the Bloch sphere.
    That is, the eigenstates of the Pauli operators, $\hat\sigma_w|w_\pm\rangle=\pm |w_{\pm}\rangle$ for $w\in\{x,y,z\}$, are the basis for our POVM,
    \begin{equation}
        \hat\Pi_{(w,\pm)}=\frac{1}{3}|w_\pm\rangle\langle w_\pm|,
    \end{equation}
    satisfying positive semi-definiteness and the completeness relation.
    For this POVM, it is easy to verify that the six measurement outcomes are
    \begin{equation}
        \vec Q
        =\frac{1}{6}
        \begin{bmatrix}
            1 & 1 & 0 & 0
            \\
            1 & -1 & 0 & 0
            \\
            1 & 0 & 1 & 0
            \\
            1 & 0 & -1 & 0
            \\
            1 & 0 & 0 & 1
            \\
            1 & 0 & 0 & -1
        \end{bmatrix}
        \begin{bmatrix}
            1 \\ x \\ y \\ z
        \end{bmatrix},
    \end{equation}
    which includes a matrix that is not sujective as the POVM is overcomplete.
    
    Since the Pauli bases are mutually unbiased, the metric tensor can be explicitly written as
    \begin{equation}
        \boldsymbol{g}
        =
        \frac{1}{18}
        \begin{bmatrix}
            2 & 0 & 1 & 1 & 1 & 1
            \\
            0 & 2 & 1 & 1 & 1 & 1
            \\
            1 & 1 & 2 & 0 & 1 & 1
            \\
            1 & 1 & 0 & 2 & 1 & 1
            \\
            1 & 1 & 1 & 1 & 2 & 0
            \\
            1 & 1 & 1 & 1 & 0 & 2
        \end{bmatrix},
    \end{equation}
    which is not invertible.
    Specifically, the nullspace can spanned by the orthogonal vectors
    \begin{equation}
        \begin{bmatrix}
            2 \\ 2 \\ -1 \\ -1 \\ -1 \\ -1
        \end{bmatrix}
        \quad\text{and}\quad
        \begin{bmatrix}
            0 \\ 0 \\ 1 \\ 1 \\ -1 \\ -1
        \end{bmatrix}.
    \end{equation}
    The four non-zero eigenvalues of $\boldsymbol{g}$ are $1/3$, for the eigenvector $\vec n_6$, and $1/9$, for the three orthogonal eigenvectors
    \begin{equation}
        \begin{bmatrix}
            1 \\ -1 \\ 0 \\ 0 \\ 0 \\ 0
        \end{bmatrix},
        \quad
        \begin{bmatrix}
            0 \\ 0 \\ 1 \\ -1 \\ 0 \\ 0
        \end{bmatrix},
        \quad\text{and}\quad
        \begin{bmatrix}
            0 \\ 0 \\ 0 \\ 0 \\ 1 \\ -1
        \end{bmatrix}.
    \end{equation}
    
    Expressing the outcome vector $\vec Q$ in the eigenbasis of $\boldsymbol{g}$ together with nullspace elements when applying the pseudo-inversion, we obtain
    \begin{equation}
    \begin{aligned}
        \vec P_\sigma
        ={}&
        \frac{9^\sigma}{6}\left(
            3^{-\sigma}
            \begin{bmatrix}
                1 \\ 1 \\ 1 \\ 1 \\ 1 \\ 1
            \end{bmatrix}
            +x
            \begin{bmatrix}
                1 \\ -1 \\ 0 \\ 0 \\ 0 \\ 0
            \end{bmatrix}
            +y
            \begin{bmatrix}
                0 \\ 0 \\ 1 \\ -1 \\ 0 \\ 0
            \end{bmatrix}
        \right.
        \\
        {}&
        \left.
            +z
            \begin{bmatrix}
                0 \\ 0 \\ 0 \\ 0 \\ 1 \\ -1
            \end{bmatrix}
            +\alpha
            \begin{bmatrix}
                2 \\ 2 \\ -1 \\ -1 \\ -1 \\ -1
            \end{bmatrix}
            +\beta\begin{bmatrix}
                0 \\ 0 \\ 1 \\ 1 \\ -1 \\ -1
            \end{bmatrix}
        \right),
    \end{aligned}
    \end{equation}
    where the parameters $\alpha$ and $\beta$ are not determined, yet.
    In order not to falsely identify the negative contributions of $\vec P_\sigma$, we select $\alpha$ and $\beta$ in such a manner that $\vec P_\sigma\geq0$ when possible.
        Requiring that the minimal entries ought to be nonnegative in that scenario results in
    \begin{equation}
    \begin{aligned}
        0\leq {}& 3^{-\sigma}- |x|+2\alpha,
        \\
        0\leq {}& 3^{-\sigma}- |y|-\alpha+\beta,
        \\
        0\leq {}& 3^{-\sigma}- |z|-\alpha-\beta.
    \end{aligned}
    \end{equation}
    The sum of the three inequalities results in the constraint
    \begin{equation}
        0\leq 3^{1-\sigma}-|x|-|y|-|z|=u
    \end{equation}
    for the existence of a solution $\vec P_\sigma \geq0$.
    Returning to our initial three inequalities for $\sigma$-classical states, we substitute $\gamma=3^{-\sigma}-|y|-\alpha+\beta$ and $\delta=3^{-\sigma}-|z|-\alpha-\beta$, meaning that the last two inequalities simplify to $0\leq \gamma$ and $0\leq \delta$.
    Solving for $\alpha=3^{-\sigma}-(|y|+|z|+\gamma+\delta)/2$ and $\beta=(|y|-|z|+\gamma-\delta)/2$, the first inequality takes the form $0\leq u-\gamma-\delta$.
    For a $\sigma$-classical state, with $u\geq0$, we simultaneously satisfy the three rewritten inequalities when selecting $\gamma=\delta=u/3$---all three taking the form $0\leq u/3$.
    From this choice that guarantees non-negative probabilities for $\sigma$-classical states, we can recast the corresponding vector of probabilities as
    \begin{equation}
        \vec P_\sigma
        =
        \frac{9^\sigma}{18}
        \begin{bmatrix}
            u+3(|x|+ x)
            \\
            u+3(|x|- x)
            \\
            u+3(|y|+ y)
            \\
            u+3(|y|- y)
            \\
            u+3(|z|+ z)
            \\
            u+3(|z|- z)
        \end{bmatrix}.
    \end{equation}
    
    The third column in Fig. \ref{fig:allPlots} shows the resulting classical states as the green, embedded set for different $\sigma$ values.
    In geometric terms, we intersect a rescaled version of octahedron that represents the POVM with the Bloch ball.
    Interestingly, the Wigner-function-like quasiprobability, $\sigma=1/2$ in the first row, does not exhibit $\sigma$-nonclassical states.
    The last row again shows a noisy version of the overcomplete measurement under study.
    
\subsection{Noisy, incomplete, and overcomplete POVM}

    Combining the observations from the previous examples, we now consider a noisy POVM that is informationally incomplete and overcomplete at the same time.
    That is, the full state cannot be reconstructed from the measurement outcomes alone, and the information for the part that can be reconstructed includes redundancies.
    This is achieved by four states forming a square in the equatorial plane of the Bloch sphere, as obtained from the POVM elements
    \begin{equation}
        \hat\Pi_{w,\pm}=\frac{1}{2}|w_\pm\rangle\langle w_\pm|,
        \quad\text{for}\quad
        w\in\{x,y\}.
    \end{equation}
    Now, we proceed as formulated in greater detail for the previous examples, while skipping analogous discussions of technical details here.
    Here, we have
    \begin{equation}
        \vec Q
        =\frac{1}{4}
        \begin{bmatrix}
            1 & 1 & 0 & 0
            \\
            1 & -1 & 0 & 0
            \\
            1 & 0 & 1 & 0
            \\
            1 & 0 & -1 & 0
        \end{bmatrix}
        \begin{bmatrix}
            1 \\ x \\ y \\ z
        \end{bmatrix}.
    \end{equation}
    The linear map involved in the above equation is neither surjective nor injective because the POVM studied now is both overcomplete and incomplete.
    The metric tensor reads
    \begin{equation}
        \boldsymbol{g}
        =
        \frac{1}{8}
        \begin{bmatrix}
            2 & 0 & 1 & 1
            \\
            0 & 2 & 1 & 1
            \\
            1 & 1 & 2 & 0
            \\
            1 & 1 & 0 & 2
        \end{bmatrix},
    \end{equation}
    whose null space is spanned by $[1,1,-1,-1]^\mathrm{T}$.
    In addition, the eigenvalue $1/2$ is obtained for $\vec n_4$, and the eigenvalue $1/4$ pertains to the eigenvectors $[1,-1,0,0]^\mathrm{T}$ and $[0,0,1,-1]^\mathrm{T}$.
    Analogously to the previous example, an optimization over the null-space contribution allows us to determine the measurement-based quasiprobabilities as
    \begin{equation}
        \vec P_\sigma
        =
        \frac{4^\sigma}{8}
        \begin{bmatrix}
            v+2(|x|+x)
            \\
            v+2(|x|-x)
            \\
            v+2(|y|+y)
            \\
            v+2(|y|-y)
        \end{bmatrix},
    \end{equation}
    with the definition and $\sigma$-classical constraint
    \begin{equation}
        0\leq 2^{1-\sigma}-|x|-|y|=v.
    \end{equation}
    
    The fourth column in Fig. \ref{fig:allPlots} shows the resulting $\sigma$-classical states as the green set.
    In geometric terms, we intersect a square-based prism with the Bloch ball which accounts for the fact that the inaccessible $z$-contribution does not alter the (non-)classicality.
    The Wigner-function-like quasiprobability, $\sigma=1/2$ in the first row, does not exhibit $\sigma$-nonclassical states;
    the forth row models a noisy version of the measurement studied in this final example.
    
\section{Conclusion}
\label{sec:Conclusion}

    We constructed and analyzed measurement-based quasiprobabilities, with the corresponding notion of nonclassicality being defined relative to the available measurements.
    This is motivated by the practical problem that commonly occurs when experimental data result in an incomplete or overcomplete set, leading to only partial or non-unique reconstructions of the quantum state.
    To determine the quasiprobabilities in such scenarios, a pseudo-inversion and optimization over null-spaces was carried out.
    Essential examples for qubit measurement demonstrate the effects that incomplete and overcomplete POVMs have on the resulting determination of nonclassical properties.

    Our methodology includes as a special case the seminal KD quasiprobabilities when weak measurements are considered in our framework.
    Moreover, partial inversion rendered it possible to generalize the notion of $ s$-parametrized phase-space representations from quantum optics to our measurement-based approach.
    With the same technique, we can model noisy measurement scenarios, enabling us to investigate and compare nonclassicality as assessed from ideal and imperfect measurement devices.
    The proof-of-concept applications we studied demonstrated the broad impact of quantum-informational completeness of available data, including noisy, incomplete, overcomplete measurement scenarios, as well as combinations thereof.

    Therefore, the proposed framework provides a versatile toolbox to define and probe notions of measurement-based nonclassicality for targeted experiments.
    Our methodology is device-agnostic and system-agnostic since it is neither restricted to any detection device nor to a specific physical system, including discrete and continuous variables, Gaussian and non-Gaussian measurements, as well as measurements in  correlated many-body systems.

\begin{acknowledgments}
    J.S. and L.A. acknowledge funding through the Quant\-ERA project QuCABOoSE.
    E.A. acknowledges funding from the Austrian Science Fund (FWF) through the Elise Richter project 10.55776/V1037.
\end{acknowledgments}

\bibliographystyle{apsrev-new}
\bibliography{bibliography}
\end{document}